\documentclass[aps,prl,reprint,amsmath,amssymb]
{revtex4-1}
\usepackage{graphicx}
\begin{document}

\title{Pressure-induced superconductivity in elemental ytterbium metal}
\author {Jing Song and James S. Schilling}
\affiliation {Department of Physics, Washington University, St. Louis, MO 63130,
USA}
\date{\today}

\begin{abstract}
Ytterbium (Yb) metal is divalent and nonmagnetic but would be expected under sufficient pressure to become trivalent and magnetic. We have carried out electrical resistivity and ac magnetic susceptibility measurements on Yb to pressures as high as 179 GPa over the temperature range 1.4 - 295 K. No evidence for magnetic order is observed. However, above 86 GPa Yb is found to become superconducting near 1.4 K with a transition temperature that increases monotonically with pressure to approximately 4.6 K at 179 GPa. Yb thus becomes the 54th known elemental superconductor.
\end{abstract}

\maketitle

The majority of elemental solids in the periodic table are 
superconductors:  31 at ambient pressure and 22 more under high pressure 
\cite{schilling10,bi}. However, of the 15 lanthanide metals, only 4 are known to superconduct:  
La at ambient pressure and Ce, Eu, Lu under high pressure \cite{schilling10}. The paucity of 
superconductivity in the lanthanides is mainly due to their 
strong local-moment magnetism. This is not true for Yb since 
its 4$f^{14}$ orbital is filled and thus nonmagnetic.

All lanthanide metals are trivalent except Eu and Yb that are
divalent.  One would anticipate, however, that under sufficient pressure both Eu and Yb would 
ultimately become fully trivalent whereby 
one 4$f$ electron would jump into the conduction band. Yb would then take on the 
magnetic $4f^{13}$ configuration and order magnetically 
at some temperature $T_{\text{o}}$. As discussed below, de Gennes factor considerations lead 
to an anticipated magnetic ordering temperature for \textit{trivalent} Yb near 6 K. 

Whereas an early x-ray absorption spectroscopy measurement found Yb to be
fully trivalent at 34 GPa \cite{syassen}, a later study over the same pressure range concluded 
that Yb's valence saturates at approximately 2.7 \cite{fuse}. Theoretical calculations indicate that
the valence of Yb does indeed increase with pressure, but the estimated degree of
increase depends on the approximations used \cite{pickett}.
The equation of state and structural phase transitions in Yb have
been determined at ambient temperature to pressures as high as 202 GPa \cite{chesnut}. These 
authors conclude that Yb is fully trivalent for pressures of 100 GPa and above.

Both Ce \cite{wittig} and Eu \cite{debessai} order magnetically but become superconducting 
under pressure when the magnetism is suppressed. Both effects may be the result of 
an increasing instability in the magnetic state under sufficient pressure. 
If so, it would seem possible that 
pressure-induced magnetic instabilities in Yb could also lead to superconductivity. This would be
of particular interest since under pressure both Ce and Eu are pushed from a magnetic to a 
nonmagnetic state, whereas in Yb exactly the opposite may occur. 
Studying the possible nonmagnetic-magnetic
transition in Yb would provide important information to further our understanding
of these highly correlated electron phenomena. For example, 
an in-depth study of elemental Yb metal would directly aid our understanding of heavy
fermion, quantum critical, non-Fermi liquid, magnetic ordering, and/or
unconventional superconductivity in Yb-based compounds 
\cite{nakatsuju,mori,cornelius,schlottmann,flouquet}, and also enhance our understanding
of these phenomena in general.
Very recently $\beta $-YbAlB$_{4}$ \cite{nakatsuju} was reported to be the first
superconducting Yb-based heavy-fermion system, with $T_{\text{c}}$ = 80 mK.

Previous high-pressure transport studies on Yb have focused on the metal-insulator 
transition below 5 GPa \cite{ramos,rice}; another resistivity experiment extended the pressure
range to 16 GPa \cite{mydosh}. In neither study were magnetic ordering 
or superconductivity observed above 2 K. 

In this letter we extend previous transport and magnetic measurements on Yb
to pressures exceeding 100 GPa (1 Mbar). Both electrical resistivity and ac magnetic susceptibility
measurements confirm that Yb becomes superconducting 
above 80 GPa at 1.4 K with $T_{\text{c}}$ increasing to $\sim$4.5 K at 179 GPa. No sign of
magnetic order is observed over this entire pressure range.

To generate
pressures surpassing 1 Mbar, a membrane-driven \cite{daniels1} diamond anvil
cell (DAC) made of CuBe alloy was used \cite{Schilling84} where pressure was
generated between two opposed diamond anvils (1/6-carat, type Ia) with 0.35
mm diameter culets beveled at 7$^{\circ }$ to 0.18 mm central flats. In the
resistivity experiment a Re gasket (6 - 7 mm diameter, 250 $\mu $m thick)
was pre-indented to 30 $\mu $m and a 90 $\mu $m diameter hole electro-spark
drilled through the center. The center section of the pre-indented gasket
surface was filled with a 4:1 cBN-epoxy mixture to insulate the gasket and
serve as pressure medium. The thin square-shaped Yb sample (Alfa Aesar, 99.9\%)
with dimensions $\sim$40$\times $40$\times $5 $\mu $m was then placed on top of four 
thin (4 $\mu $m) Pt leads for a four-point dc electrical resistivity measurement with
0.5 mA excitation current. See paper of Shimizu \textit{et al.} \cite{Shimizu05} 
for further details of the non-hydrostatic pressure technique.
 
\begin{figure}[t]
\includegraphics[width = 8 cm]{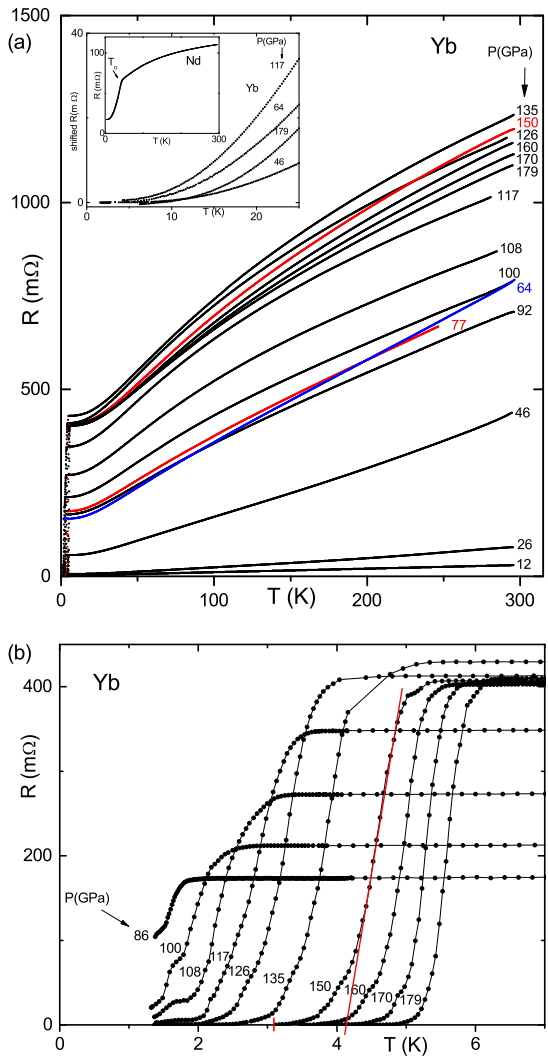} 
\caption{\label{fig1}(color online) (a) Resistance of Yb in run R1 versus temperature from
1.4 K to 295 K at room-temperature pressures to 179 GPa. Inset shows smooth $R(T)$ 
dependence below 25 K; note that residual resistance at 5 K has been subtracted off. 
Inset inside inset shows sharp break in slope of $R(T)$
 for Nd signaling magnetic ordering at $T_{\text{o}}$ $\simeq $ 44 K \cite{song1}.
 (b) Resistance data from (a) for temperatures 1.4 K to 7 K showing
superconducting transitions for pressures 86 GPa and above.
Data at 86 GPa are from run R2. Here $T_{\text{c}}$ is defined (see, for example, data at 150 GPa) 
as temperature where straight red line through data hits temperature axis; vertical tick marks
temperature where $R(T)$ vanishes.}
\end{figure}
 
Since a Re gasket superconducts under pressure at $\sim $4 K, a gasket made of 
MP35N (neither superconducting nor magnetic) was used in the ac susceptibility $\chi (T)$
measurements. The MP35N gasket was heat treated at 565$^{\text{o}}$C for 4
hours resulting in an increase in HRC hardness from 48 to 52. The 90 $%
\mu $m hole in the pre-indented gasket was filled with Yb sample without
pressure medium. The real and
imaginary parts of $\chi (T)$ were determined in the first harmonic, and at
some pressures in the third harmonic, using a Stanford Research SR830 digital
lock-in amplifier with a SR554 transformer preamplifier and a Keithley 6221
constant ac current source.

In all experiments the pressure at room temperature was determined by Raman spectroscopy from 
the diamond vibron \cite{raman1}. In selected experiments on Yb a ruby manometer \cite{Chijioke05} 
was also used, revealing an approximately linear increase in pressure 
of about 20\% on cooling from 295 to 4 K. Since the important phenomena either occur or are expected to occur at temperatures near 4 K, all values of pressure in this paper are enhanced by 20\% over those measured at ambient temperature using the diamond vibron. Further experimental details of the DAC, cryostat, and ac
susceptibility techniques are given elsewhere \cite{Schilling84,klotz1,debessai1}.

The temperature dependence of the
resistivity for Yb, $R(T,P)$, was determined for pressures exceeding 1 Mbar in
two separate experiments. The data obtained in the first are
shown in Fig 1(a) and its inset for pressures to 179 GPa. Careful examination reveals that all $R(T)$ data increase smoothly 
with temperature over the entire temperature range 1.4 - 295 K, giving no evidence for magnetic order. 
The inset displays for an earlier experiment on Nd \cite{song1} the typical break in 
slope of $R(T)$ or resistivity knee that signals the
occurrence of magnetic ordering at the temperature $T_{\text{o}}$ = 44 K. 
At both ambient and low temperatures the
resistance of Yb is seen to increase with pressure to 135 GPa, but then
to decrease at higher pressures. This may result from the fcc-hP3 phase
transition.

For pressures of 86 GPa and above the resistance at low temperatures is seen
to fall towards zero, pointing to a transition to superconductivity. This is illustrated
more clearly in Fig 1(b) where the resistance is seen to fall completely to zero at
most pressures to 179 GPa. In addition, no change in the shape
of the transition is observed if the current is reduced form 0.5 mA to 0.1 mA, thus pointing
to bulk, rather than filamentary, superconductivity. These results were 
confirmed by a second resistivity experiment to extreme pressures.

A superior test for superconductivity is a measurement of the magnetic susceptibility.
In Fig 2(a) the real part of the temperature-dependent ac magnetic
susceptibility $\chi \prime (T)$ of Yb is shown for one of three experiments. The large
negative magnetic shielding effect seen is consistent with full screening from bulk
superconductivity at 109, 116, 123, and 132 GPa, whereas no evidence for
superconductivity is seen above 1.4 K at 92 GPa. To enhance the visibility 
of the superconducting transition, the
temperature-dependent background signal at 72 GPa has been subtracted 
from the data shown. The inset to Fig
2(a) shows the raw data for $\chi \prime (T)$ at 116 GPa. The
superconducting transition is found to shift to lower temperatures under
magnetic fields to 250 Oe at the rate 1.36 mK/Oe. 

\begin{figure}[t]
\includegraphics[width = 7 cm]{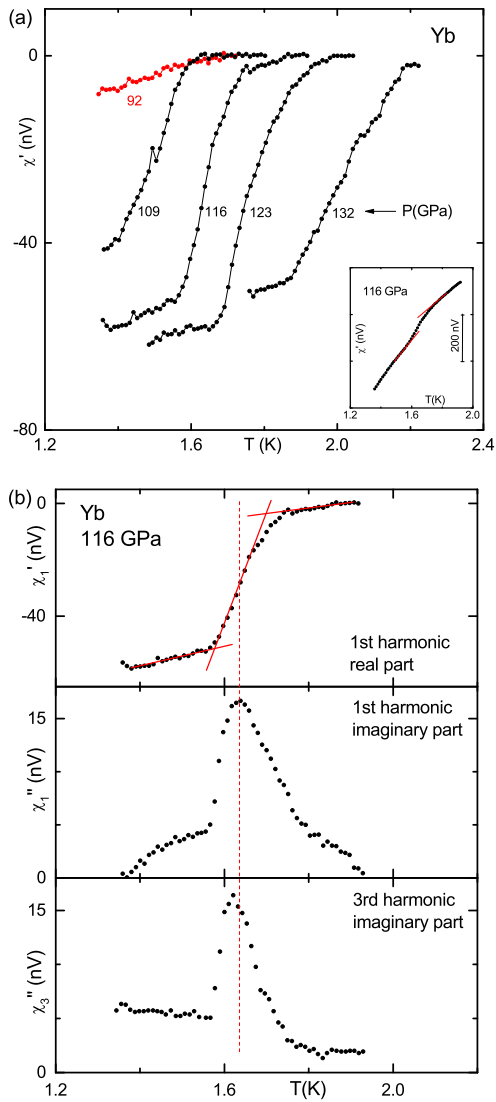} 
\caption{\label{fig2}(color online) (a) Real part of 1st harmonic of
ac susceptibility versus temperature for Yb in run X3 showing strong
diamagnetic shielding of superconducting transition for room-temperature pressures above
92 GPa. Background signal from non-superconducting Yb at 72 GPa
has been subtracted from data. Inset shows superconducting
transition at 116 GPa in raw data. (b) For Yb at 116 GPa, temperature
dependence of 1st harmonic of real and imaginary parts of ac
susceptibility with background was subtracted as in (a) and compared
to 3rd harmonic of imaginary part of ac susceptibility where no
background subtraction was necessary. All three susceptibilities clearly
define superconducting transition temperature.}
\end{figure}

\begin{figure}[t]
\includegraphics[width = 7 cm]{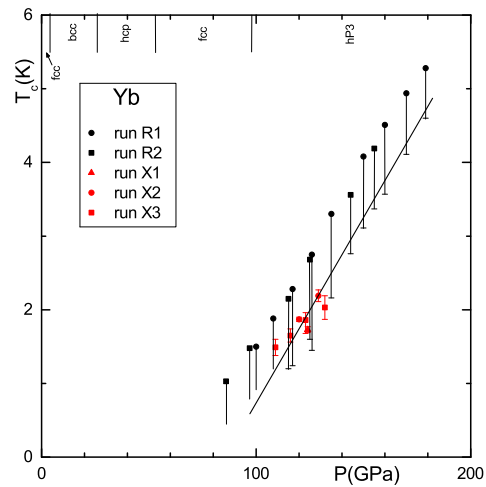} 
\caption{\label{fig3}(color online) Superconducting transition
temperature of Yb versus pressure for all resistivity and ac
susceptibility measurements. In the legend \lq\lq{R}\rq\rq and \lq\lq{X}\rq\rq are 
values from resistivity and ac susceptibility measurements, respectively. 
For resistivity $T_{\text{c}}$ is defined as
temperature where $R(T)$ extrapolates to zero, upper error bar
marking transition midpoint and lower error bar where $R(T)$
actually reaches zero. In ac susceptibility $T_{\text{c}}$ is defined as
temperature at transition midpoint for real part of 1st
harmonic, upper and lower error bars giving temperatures where straight
line through data intersect, as in upper panel of Fig. 2(b). Long straight line gives 
best estimate of $T_{\text{c}}$ versus pressure with slope 50 mK/GPa. Structures at top of
graph for Yb follow the structure sequence at room temperature \cite{chesnut}:  
fcc(I) to bcc at 4 GPa, to hcp at 26 GPa, to fcc(II) at 53 GPa, to hP3 at 96 GPa.}
\end{figure}

The imaginary part $\chi_{1}\prime \prime (T)$ in the 1st harmonic was also measured at all
pressures as well as the imaginary part in the 3rd harmonic $\chi _{3}\prime
\prime ;$ all three are compared in Fig 2(b). Whereas the
temperature-dependent background signal was subtracted for the two 1st
harmonic susceptibilities, no subtraction was necessary for $\chi _{3}\prime
\prime $. We define $T_{\text{c}}$ to be the temperature of the midpoint of the 
transition in $\chi_{1}\prime (T)$, a temperature that corresponds approximately
to that where the resistivity falls to zero \cite{lortz1}. These results were confirmed by two further 
ac susceptibility experiments. 

During the entire resistivity experiment no electrical contact occurred between the
sample and Pt contact strips with the metal gasket or pressure cell. To
check whether the Pt contact strips themselves might become superconducting
at  extreme pressure, a separate experiment was carried out on a Pt
sample alone. No evidence for a superconducting transition was seen in the
resistivity measurements to 168 GPa pressure above 1.4 K. To check whether the MP35N
gaskets used in the ac susceptibility measurements might become
superconducting under pressure, a separate experiment was carried out on an empty gasket
containing no Yb sample. Again, no sign of a superconducting signal was observed. The
present experiments thus show that Yb metal indeed becomes superconducting
for pressures of 86 GPa and above.

In Fig 3 the values for the superconducting transition temperature 
$T_{\text{c}}$ from both resistivity and ac susceptibility measurements are plotted
versus pressure. \ The larger pressure range for the resistivity is
due to differences in pressure techniques. Within experimental error the 
resistivity and ac susceptibility measurements agree and find that under pressure 
$T_{\text{c}}$ for Yb increases at the rate of approximately +50 mK/GPa.

The overriding effect of high pressure on matter is to turn insulators into metals, quench magnetism, and promote superconductivity. The magnetism of the lanthanides with their highly localized 4\textit{f} orbitals is particularly resistant to pressure quenching. For most lanthanides Mbar pressures only suffice to generate the approach to a reduction in the number of electrons in the 4\textit{f} orbital \cite{schilling10,song1}. In the case of the highly compressible divalent non-magnetic lanthanide Yb, a full reduction would seem more likely, leaving a magnetic 4${f^{13}} $ state. X-ray spectroscopy studies differ on whether or not Yb becomes fully trivalent under 40 GPa pressure \cite{syassen,fuse}. Equation of state studies have been interpreted to support trivalent Yb at 100 GPa pressure and above \cite{chesnut}. Should this be true, magnetic order in Yb would be expected.

If one applies simple de Gennes scaling \cite{blundell} to estimate the magnetic ordering temperature of trivalent Yb, one finds that it should be approximately 49-times lower than that of Gd or $T_{\text{o}}$ = (292 K)/49 K = 6 K. Using the ratio of the magnetic ordering temperatures of GdRh$_6$B$_4$ and YbRh$_6$B$_4$ \cite{pontkees}, instead of the de Gennes factor, one arrives at the estimate $T_{\text{o}}$ = 4 K. In any case the present studies find no evidence for magnetic order in Yb above 1.4 K to pressures as high as 179 GPa. This, plus the fact that Yb becomes superconducting, speaks strongly against a fully trivalent state in Yb to this pressure. 

Although it is certainly possible that Yb is a BCS superconductor, the fact that a nonmagnetic-to-magnetic transition is approached under pressure makes an exotic form of superconductivity seem possible where magnetic fluctuations play an important role \cite{lonzarich}. This scenario has been proposed for the recently discovered heavy Fermion superconductor $\beta $-YbAlB$_{4}$ that happens to be positioned very near to a quantum critical point \cite{nakatsuju}. In typical Kondo-lattice systems the Kondo temperature and the magnitude of the negative covalent mixing exchange $J$ increase with pressure, passing through a Doniach-like phase diagram until magnetism is quenched at a quantum critical point \cite{schilling10,song1}, near where superconductivity may appear. In contrast, in Yb one would expect everything to go in reverse where the Kondo temperature decreases with pressure as the magnetic state stabilizes. Thus one might conjecture that the superconducting transition temperature will pass through a maximum and decrease before magnetic order sets in as Yb passes through the quantum critical point in reverse. To access these phenomena, the application of multi-megabar pressures would likely be necessary.

In summary, in the present resistivity and magnetic susceptibility experiments superconductivity has been discovered in Yb for pressures above 86 GPa with no sign of magnetic order between 1.4 and 295 K to 179 GPa. To this pressure Yb is clearly not in a fully trivalent, stable magnetic state. Yb thus becomes the 54th elemental superconductor discovered to date. 

\noindent \textbf{Acknowledgments.} The authors would like to thank G. Fabbris, A. Gangopadhyay, and D. Haskel for critically reading the manuscript. This work was supported by the National
Science Foundation (NSF) through Grant No. DMR-1104742 and No. DMR-1505345
as well as by the Carnegie/DOE Alliance Center (CDAC) through NNSA/DOE Grant
No. DE-FC52-08NA28554. \label{biblio}

\end{document}